\documentclass[prl,showpacs,aps,twocolumn,nofootinbib,superscriptaddress]{revtex4}
\usepackage{graphicx,amssymb,amsmath}

\begin{document}

\title{Bimodal energy relaxation in
quasi-one-dimensional compounds \\
with a commensurate modulated ground state}

\author{J. C. Lasjaunias}

\affiliation{Centre de Recherches sur les Tr\`{e}s Basses Temp\'{e}ratures, 
CNRS, BP 166, 38042 Grenoble cedex 9, France}

\author{R. M\'elin}

\affiliation{Centre de Recherches sur les Tr\`{e}s Basses Temp\'{e}ratures, 
CNRS, BP 166, 38042 Grenoble cedex 9, France}

\author{D. Stare\v{s}ini\'{c}}

\affiliation{Institute of Physics, Hr-10 001 Zagreb, P.O.B. 304, Croatia}

\author{K. Biljakovi\'{c}}

\affiliation{Institute of Physics, Hr-10 001 Zagreb, P.O.B. 304, Croatia}

\author{J. Souletie\cite{deces}}

\affiliation{Centre de Recherches sur les Tr\`{e}s Basses Temp\'{e}ratures, 
CNRS, BP 166, 38042 Grenoble cedex 9, France}

\begin{abstract}

We show that the low temperature ($T<0.5$~K) time dependent non-exponential
energy relaxation of quasi-one-dimensional (quasi-1D) compounds strongly differ
according to the nature of their modulated ground state. For 
incommensurate ground states, such as in (TMTSF)$_2$PF$_6$ the relaxation
time distribution is homogeneously shifted to larger time when the
duration of the heat input is increased, and exhibits in addition a scaling
between the width and the position of the peak in the relaxation time
distribution, 
$w^{2}\sim\ln{(\tau_{m})}$. For a commensurate ground state, as in
(TMTTF)$_2$PF$_6$, the relaxation time spectra show a bimodal
character with a weight transfer between well separated slow and
fast entities. Our interpretation is based on the dynamics of
defects in the modulated structure, which depend crucially
on the degree of commensurability.
\end{abstract}

\date{\today}
\maketitle

Non exponential relaxation is the fingerprint of disorder in a system
subject to 
an external perturbation \cite{Philips96}.
It is the consequence of a broad 
distribution of relaxation times.
The question of determining the
nature and spatial extent of correlated objects responsible
for slow dynamics in spin glasses and glasses, has been debated
for several decades.
In spin glasses,
the memory and chaos experiments \cite{Jonason98}
give information about excitations in real space.
In glasses
there is a consensus that structural relaxation
occurs on correlated domains of a nanometer size~\cite{Richter02}.
In colloidal suspensions, a direct three dimensional (3D) imaging of 
the cooperative motion in structural relaxation has been
obtained on micrometer scales~\cite{Weeks}.
A general framework for describing spatial and temporal
heterogeneities in aging systems has been proposed in Ref. \cite{Castillo}.
A new class of mesoscopic glass, with characteristic scales of
micrometers, has been recognized among quasi-one-dimensional
(quasi 1D) systems~\cite{Biljak89}.
Heat relaxation at very low temperature $T$,
below $0.5$~K, in these systems, has been shown to be non exponential
and to exhibit aging~\cite{Biljak89} in which the
entire wide spectrum of relaxation times is shifted to longer times with
increasing the duration of the very small heat perturbation.
We suggest here on the basis of heat relaxation experiments
that the formation of correlated objects in modulated
quasi-1D systems,
is qualitatively
different in commensurate and incommensurate systems.

Quasi 1D systems exhibit a variety of modulated electronic ground states at
low-temperature both in the charge channel, such as charge density wave (CDW),
and in the spin channel, such as spin 
density wave (SDW), spin-Peierls (SP) or antiferromagnetic ground state. 
Extensive investigations of the low-$T$ thermodynamical 
properties \cite{Biljak89,Biljak91,Lasjau96,Lasjau02} have shown that 
all of these systems demonstrate well-known universal features of 
glasses regardless the actual ground state. This involves an 
extra contribution to the specific heat $C_{p}$ coming from low energy excitations, and non exponential heat relaxation dynamics for $T<0.5$ K
($T$-range dominated by these excitations). 
``Aging'' effects are also involved, where 
heat relaxation and correspondingly $C_{p}$ depend on the duration of a small 
heat perturbation (a few mK)~\cite{Biljak89,Biljak91,Lasjau96}.
Extremely long relaxation times (up to $10^4$ s)
show that these excitations are only weakly coupled to the phonon
heat-bath~\cite{Lasjau96},
contrary to conventional glasses.

The origin of these excitations can be inferred
from the low frequency dynamics of 
collective excitations in the charge channel. Low frequency dielectric 
response of CDW \cite{Star04}, SDW \cite{Lasjau94} which 
also has CDW component \cite{Pouget96}, all 
having a modulated charge density, is characterized by an overdamped 
(relaxational) mode slowing down with decreasing temperature. This has 
been recognized in the case of CDW and SDW (DW) as the sign of a glass 
transition at the level of electronic superstructure. Freezing is naturally 
generated by screening effects and occurs if there are not enough free 
carriers to screen efficiently elastic DW phase deformations. Consequently, 
the characteristic scale of a DW glass is given by the phase coherence length,
which is of the order of a $\mu$m.
Below the 
glass transition temperature $T_{g}$ the low frequency dynamics
and therefore the low energy excitations
are governed by the remaining degrees of freedom -- topological or plastic 
deformations such as solitons, domain walls or dislocation loops. 

In this Letter we contrast two isostructural systems, (TMTSF)$_{2}$PF$_{6}$, 
which has an incommensurate mixed SDW-CDW ground state below 12 K,  and 
(TMTTF)$_{2}$PF$_{6}$, which exhibits a
charge order transition with a ferroelectric behavior 
at 70 K \cite{Monceau01}, and  a spin-Peierls
transition at 12 K, both 
being commensurate to the underlying lattice (in the following we use 
the notation SDW-PF$_{6}$ and SP-PF$_{6}$ respectively). We will show 
how the degree of commensurability affects the very low-$T$ dynamics and 
discuss these findings in view of the properties of
solitonic excitations.

\begin{figure}[tbp]
\centerline{\includegraphics[width=8cm]{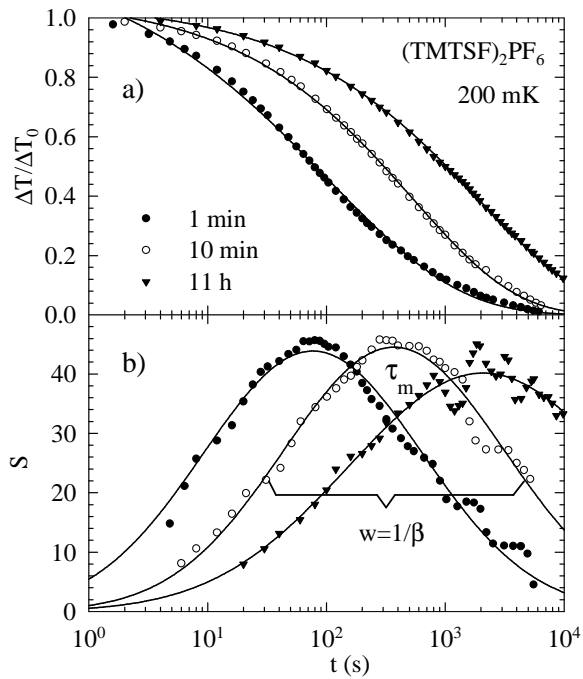}}
\caption{\textit{Homogeneous} broadening and shift
of the heat relaxation with 
increasing $t_{p}$ (aging)
for SDW-PF$_{6}$. a) Relaxation $\Delta T/\Delta T_{0}$ 
vs. $\log{t}$ with corresponding stretched exponential fit
$\exp{(-t/\tau)^\beta}$, 
b) $S(t)=d(\Delta T/\Delta T_{0})/d(\ln{t})$ fitted to a Gaussian 
function $G(\ln{t})$ with a width $w$. The direct estimate of width at half
height shown in b)
is $2.17$ decades in agreement with $1/\beta=2.15$ and $w=2.1$.}
\label{fig1}
\end{figure}
We have measured the low-$T$ heat capacity by standard relaxation method. 
The system is first stabilized at a temperature $T_{0}$. Then a small 
power is fed to the sample during a pumping time $t_{p}$.
The temperature is increased to the new, slightly higher value
$T_0+\Delta T$ ($\Delta T / T_0 < 0.05$). At $t=t_p$ the power is switched off
and heat relaxation from the sample to the cold source towards equilibrium
at $T_0$ is recorded as a function of time.
For all the CDW and SDW systems 
studied previously, below 1~K relaxation starts to deviate from 
exponential on lowering $T$ and also exhibits a strong dependence on $t_p$ 
\cite{Biljak89,Biljak91} (Fig. \ref{fig1}) . A better characterization of 
non exponential relaxation can be obtained by the relaxation rate
$S(t)=d(\Delta T/\Delta T_{0})/d(\ln t)$.
$\Delta T(t)/ \Delta T_0$ can be written in terms of a distribution
of relaxation times $P(\ln{\tau})$ as
\begin{equation}
\frac{\Delta T(t)}{\Delta T_0} = \int_{\ln{\tau_0}}^{\infty} 
P(\ln{\tau}) \exp{(-t/\tau)} d \ln{\tau}
,
\end{equation}
where $\tau_0$ is the microscopic time. Since $P(\ln{\tau})$ varies
slowly with $\tau$ we replace the exponential
by a step function, from what we deduce $S(t)=P(\ln{t})$~\cite{Lundgren}.
Like in spin glasses there is a
``homogeneous'' broadening and shifting 
of the relaxation peak
with $t_{p}$ (see Fig.~\ref{fig1}-(b)).
However, since
our system eventually reaches thermodynamic equilibrium
at $T_0+\Delta T$ for sufficiently long $t_p$~\cite{Biljak91},
aging is then interrupted~\cite{Bouchaud}.

\begin{figure}[tbp]
\centerline{\includegraphics[width=8cm]{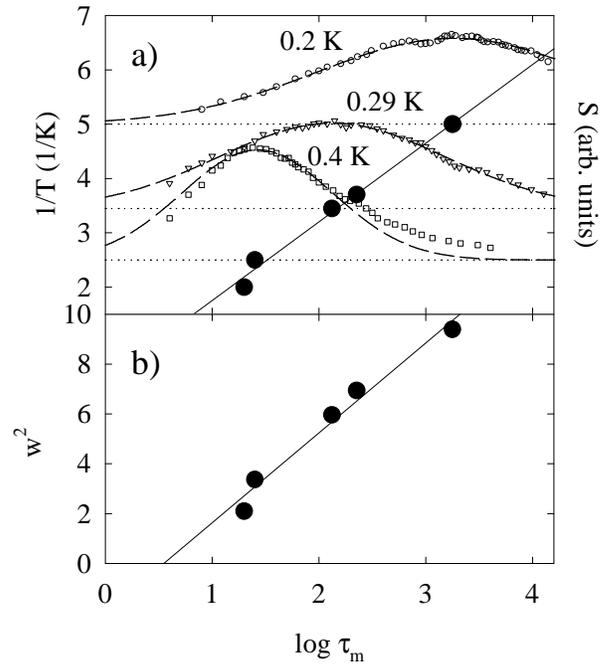}}
\caption{a) An ``inverted'' Arrhenius plot, $1/T$ vs. $\log{\tau_{m}}$, 
obtained for relaxations in thermodynamical equilibrium ($\tau_{m}=\tau_{eq}$) 
of SDW-PF$_{6}$. Relaxation rates and their fits are only shown for three selected 
temperatures among five.
b) The $w^{2}\sim \ln{\tau_{m}}$ scaling is well obeyed as expected 
from central limit theorem. The activation energy is $E_a=1.7$~K.}
\label{fig2}
\end{figure}

Non exponential relaxation can be fitted by a ``stretched exponential'' 
$\Delta T/\Delta T_{0} \sim \exp{[-(t/\tau)^{\beta}]}$ \cite{Biljak89}, 
while the relaxation rate $S(t)$ can be fitted by a Gaussian distribution in 
$\ln{t}$: $G(\ln{t})=(1/\sqrt{2\pi w}) \exp{(-\ln^{2}{(t/\tau_{m})/2 w^{2}})}$, 
centered on $\ln{\tau_{m}}$ and with a width $w$. Indeed, the logarithmic 
derivative of a stretched exponential $d(\exp{(-(t/\tau)^\beta)})/d(\ln{t})$ is almost
 a Gaussian with $\tau_m\sim\tau$ and $1/w\sim \beta$, as shown in
Fig.~\ref{fig1}. 
The same $w$ can also be extracted directly from $S(t)$ curves as the width at half height. 

While the stretched exponential has no known justification other than 
phenomenological \cite{Philips96,Richter02}, Castaing and Souletie \cite{Castaing91} 
argued that a Gaussian distribution in $\ln{t}$ can be justified in a framework 
that associates a hierarchical criterion and a usual scaling approach in the 
renormalization of distances. Similarly to turbulence they use a cascade of events 
to relate the macroscopic and microscopic scales. After iterating the renormalization 
group dynamical scaling takes the form  $\tau/\tau_{0}\sim (\xi/\xi_{0})^{z}$, 
where $z$ is the dynamical exponent. The time scale distributions are log-normal: 
$P(\ln{\tau})\equiv G(\ln{t})$ is Gaussian. The variance $w^{2}$ scales with the 
most probable value $\ln{(t/\tau_{m})}$ \cite{height} following the central limit 
theorem.  It is 
demonstrated in Fig. \ref{fig2} for SDW-PF$_{6}$ in thermodynamical equilibrium 
\cite{Lasjau02} for three temperatures chosen below $0.5$ K. This yields immediately 
the temperature dependence of the stretched exponential exponent $\beta(T)\sim\sqrt{T}$, as 
$\beta\sim 1/w\sim 1/\sqrt{\ln{\tau_{m}}}$ (Fig. \ref{fig2}b), so that $\ln{\tau_{m}}\sim 1/T$
indicates a thermally activated process with activation energy $E_{a}=1.7$ K. At this
 point we do not intend to go deeper into the different models related to $\beta(T)$ for
$T\rightarrow 0$ (see ref. \onlinecite{Biljak89} and references therein). We stress that 
the $w^{2}\sim \ln{\tau_{m}}$ scaling is quite convincing in the range of interest, 
and also for TaS$_{3}$, a prototype CDW system, for which as well as for SDW-PF$_{6}$ 
a glass transition has been found at higher $T$ \cite{Star04}. 

\begin{figure}[tbp]
\centerline{\includegraphics[width=8cm]{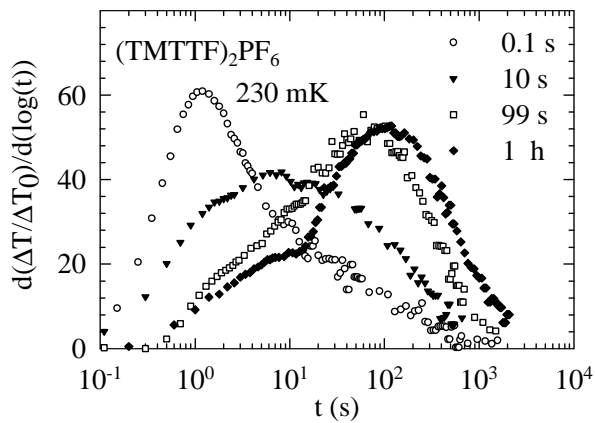}}
\caption{\textit{Bimodal} relaxation with redistribution of the
spectral weight
and saturation for $t_{eq}\sim 10^{2}$ s for
the spin-Peierls compound SP-PF$_{6}$. 
The values of $t_p$ 
 are indicated on the figure}
\label{fig3}
\end{figure}

However, in the case of SP-PF$_{6}$ \cite{Lasjau02} heat relaxation, while still 
non exponential, exhibits quite different properties. Instead of the
homogeneous broadening and shift
of the spectra of SDW-PF$_{6}$ as in a 
real ``aging'' effects, 
SP-PF$_{6}$ demonstrates ``discrete bands'' of relaxation times. Their 
distributions are narrower and the
relaxation times are smaller than in SDW-PF$_6$ (i.e. saturation 
of aging is reached within a few tens of
minutes as compared to days in first system at a similar $T$).  
Fig. \ref{fig3} shows a bimodal redistribution of the relaxation spectrum indicating that
 heat perturbation modifies different parts of the spectrum in contrast to the
homogeneous shift of the entire spectrum as in Fig.~\ref{fig1}.

\begin{figure}[tbp]
\centerline{\includegraphics[width=8cm]{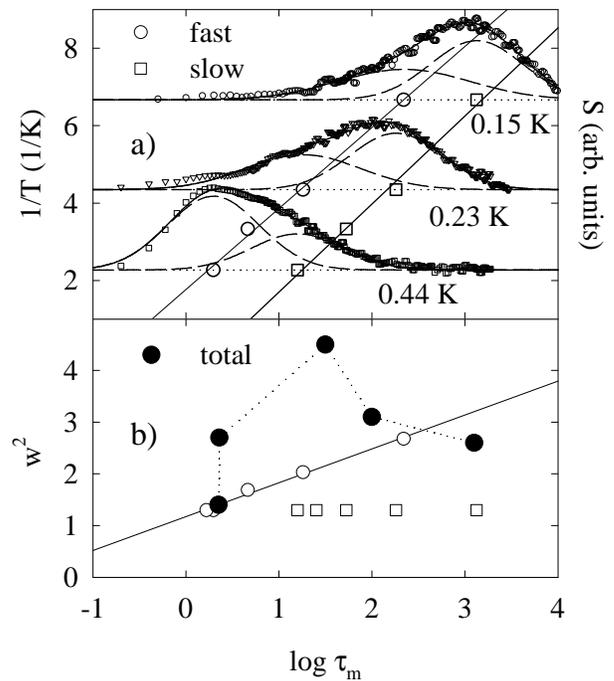}}
\caption{a) An ``inverted'' Arrhenius plot, $1/T$ vs. log$\tau_{m}$, obtained for the 
relaxations in thermodynamical equilibrium of
the spin-Peierls compound SP-PF$_{6}$, together with a representation 
of corresponding relaxation rates for slow and fast processes and their fits, with 
$E_{a,s}= E_{a,f}=1.0$~K b)  The $w^{2}\sim \ln{\tau_{m}}$  scaling does not work for 
the total relaxation, however it might work for each deconvoluted process.}
\label{fig4}
\end{figure}

Even in thermodynamical equilibrium the $S(t)$ curves are not symmetric,
indicating at least two different relaxation processes. However the total $w$
(Fig.~\ref{fig4}) can be obtained, but the variation $w^2
\sim \ln{\tau_m}$ is very peculiar. We have successfully deconvoluted
$S(t)$ in two dynamically distinct Gaussian contributions, 
represented by dashed lines in Fig. \ref{fig4}a. At higher $T$, the fast contribution 
plays the dominant role in the dynamics, but the relative weight of slow modes in the 
dynamics grows on decreasing the temperature. A similar behavior occurs at a fixed 
temperature if we change the duration $t_p$ of the heat perturbation (Fig. \ref{fig3}). 
There is an evident weight transfer from one to the other subsystem as thermodynamic 
equilibrium is approached. However,  without the fast entities, there would be almost no 
response within $10^{2}$-$10^{3}$ seconds at the lowest $T$. The 
distribution of fast modes is pushed to $10^{3}$ seconds at 85 mK,
and the apparent activated behavior (Fig. \ref{fig4}a) turns into a saturation indicating 
possible tunneling effects (not included in Fig. \ref{fig4}). 

There are drastic consequences on the proper definition and determination of the specific 
heat $C_{p}$ in these systems as it becomes strongly nonlinear \cite{Lasjau96}. From the 
deconvoluted entities in thermodynamic equilibrium (Fig. \ref{fig4}a), it is possible to 
calculate the corresponding $C_{p}$, fast and slow. Their participation changes 
progressively from approximately 100 \% for the fast modes at $\sim 650$ mK to almost 
90\% for the slow modes below 400 mK.

It appears that the incommensurate CDW compound
o-TaS$_3$ shares all its properties related to the shift of the
dynamics
with SDW-PF$_{6}$. Both are nominally {\underline {incommensurate}} systems. On the other hand, 
(TMTTF)$_{2}$Br shares its multimodal dynamics properties with SP-PF$_{6}$ 
\cite{Lasjau02}. (TMTTF)$_{2}$Br being the brother compound of SP-PF$_{6}$ is a 
\underline{commensurate} antiferromagnet.
We conclude our Letter by proposing an explanation to these
distinctly different
dynamics. The models generalizing Refs.\cite{Larkin,Ov,revue} are investigated in more details 
in Ref. \cite{Melin02}.

As proposed by Larkin~\cite{Larkin} and Ovchinnikov~\cite{Ov},
the local deformations of the DW in the strong pinning limit
are bisolitons,
obtained by minimizing the elastic and pinning energies,
and characterized by a ground state at energy $E_0$, separated from
another state at energy $E_0+\Delta E$ by a ``bounce'' state at energy
$E_0+\Delta V$,
therefore defining an effective two-level system.
As a qualitative difference between commensurate and incommensurate
systems\cite{Melin02},
one has $\Delta E=0$
in the commensurate case~\cite{note} and $\Delta E >0$ in the incommensurate
case, so that the populations of the effective two level system does 
not couple to temperature variations in the commensurate case.
We suggest here that, in addition, the degeneracy of the classical
ground state in the commensurate 
case is lifted by quantum fluctuations that restore a 
finite heat response.
The long time dynamics in the incommensurate case is due to
collective effects, as suggested
by the $w^2 \sim \ln{\tau_m}$ scaling on Fig.~\ref{fig1},
in agreement with Ref.~\cite{Melin02}.
The issue of the long time dynamics in the
commensurate case, involving both quantum and collective effects,
is mainly an open theoretical question.

To explain why the spectrum of relaxation 
time is bimodal in the commensurate case we note that
experimental data suggest that in the commensurate system there exist two 
entities that relax with approximately the same energy barrier but with values of the 
``microscopic time'' differing by one order of magnitude. Possible candidates for these two 
entities would be (i) the dipole solitonic excitations generated by the strong pinning 
impurities; and (ii) self-induced disorder due to $2 \pi$-solitons
between micro-domains separating 
the two spin-Peierls ground states. 
Alternatively one may observe that SP-PF$_6$ has a 
charge gap $\Delta_C \simeq 200$ K coexisting
with a spin gap $\Delta_S \simeq 20$ K, deduced from the spin-Peierls
transition temperature from the
mean field approximation. 
The coherence length in the charge channel $\xi_C=\hbar v_F/\Delta_C \simeq 10 a_0$ is thus
much smaller than the coherence length in the spin channel
$\xi_S=\hbar v_F/\Delta_S \simeq 100 a_0$ (where $a_0$ is the lattice
parameter and $v_F$ the Fermi velocity).
 The two channels are weakly coupled so that
the charge soliton, being smaller,  relaxes faster than the spin soliton.
The pinning potential and thus the energy barriers are identical in the two sectors
so that the activation energies are identical but the microscopic times are
different, in agreement with Fig.~\ref{fig4}.

In conclusion we have shown manifestations of correlated objects interpreted 
as the collective dynamics of a disordered soliton lattice inherent to the modulated 
electronic ground state of DW. We demonstrated that a simple scaling feature of the 
parameters of the relaxation time distribution is well obeyed for the homogeneously 
broadened response for incommensurate systems.
By contrast the heat response of commensurate systems is bimodal.
We stress that our results should be also 
compared to other experiments  on the dynamical
heterogeneity in the relaxations of supercooled liquids or spin glasses
\cite{Schiener,Chamberlin,Jeffrey}, and to the case of
bimodal relaxation \cite{Sen97}
observed in a fragile glass former at the approach of a glass transition.
However, the microscopic picture is different, in spite of
a similar phenomenology.

\end{document}